\documentclass{article}

\usepackage{ifthen}
\usepackage{url}  %
\usepackage{siunitx}
\usepackage{etoolbox}
\usepackage{booktabs}
\usepackage{xspace}
\usepackage{amsmath}

\newboolean{bioinfo}
\setboolean{bioinfo}{false}

\ifthenelse{\boolean{bioinfo}}{}{
\usepackage[letterpaper,margin=1in]{geometry}
}

\makeatletter
\ifthenelse{\boolean{bioinfo}}{
               {\list{}{\labelwidth\z@ \itemindent-\leftmargin
                        }}
               {\endlist}

\setlength{\labelsep}{.5em}%
}{}
\makeatother

\usepackage{mdwlist}

\RequirePackage{natbib}

\DeclareMathOperator{\cnt}{Cnt}

\begin{document}

\ifthenelse{\boolean{bioinfo}}{}{
\newcommand{\address}[1]{\date{#1 \\ \bigskip \today}}
\newcommand{\firstpage}[1]{\setcounter{page}{#1}}
\newcommand{\history}[1]{}
\newcommand{\editor}[1]{}
\newcommand{\href}[2]{#2}
\newenvironment{methods}{}{}
\newcommand{\processtable}[3]{\caption{#1}\center #2}
\def\textcolon{\text{\rm :}}

\let\oldtitle\title
\renewcommand{\title}[2][]{\oldtitle{#2}}
\let\oldauthor\author
\renewcommand{\author}[2][]{\oldauthor{#2}}

\newcommand{\absection}[1]{\par\noindent #1\space\ignorespaces}
\renewenvironment{abstract}{%
  \begingroup
  \let\section\absection
  {\center ABSTRACT}\par}
{\endgroup\bigskip}
}

\firstpage{1}

\newcommand{\quorum}{QuorUM\xspace}
\hyphenation{quor-um}
\newcommand{\species}[1]{\textit{#1}}
\newcommand{\cov}[1]{\ensuremath #1\textrm{x}\xspace}
\newcommand{\p}[1]{\textbf{#1}}
\newcommand{\winner}[1]{\bfseries #1}

\title[\quorum]{\quorum: an error corrector for Illumina reads}
\author[Mar\c{c}ais \textit{et~al}]{Guillaume Mar\c{c}ais\,$^{1,*}$, James A. Yorke\,$^{1}$ and Aleksey Zimin\,$^1$\footnote{to whom correspondence should be addressed}}
\address{$^{1}$University of Maryland, College Park, MD}

\history{Received on XXXXX; revised on XXXXX; accepted on XXXXX}

\editor{Associate Editor: XXXXXXX}

\maketitle

\begin{abstract}

\section{Motivation:}
Illumina Sequencing data can provide high coverage of a genome by relatively short (\SI{100}{bp} to \SI{150}{bp}) reads at a low cost.
Our goal is to produce trimmed and error-corrected reads to improve genome assemblies.
Our error correction procedure aims at producing a set of error-corrected reads (1) minimizing the number of distinct false $k$-mers, i.e.\@ that are not present in the genome, in the set of reads and (2) maximizing the number that are true, i.e.\@ that are present in the genome.
Because coverage of a genome by Illumina reads varies greatly from point to point, we cannot simply eliminate $k$-mers that occur rarely.

\section{Results:}
Our software, called \quorum, provides reasonably accurate correction and is suitable for large data sets ($1$ billion bases checked and corrected per day per core).

\section{Availability:}
\quorum is distributed as an independent software package and as a module of the MaSuRCA assembly software. Both are available under the GPL open source license at \url{http://www.genome.umd.edu}.

\section{Contact:} \href{gmarcais@umd.edu}{gmarcais@umd.edu}
\end{abstract}

\section{Introduction}
While second generation sequencing technologies have progressed tremendously and offer ever longer reads with low overall sequencing error rate, correcting errors in reads remains an important pre-processing step in \emph{de novo} genome assembly.
Most current assembly software use the de~Bruijn graph representation as one step of the assembly process~\citep{zerbino_velvet:_2008,li_novo_2010,chaisson_short_2008,gnerre_high-quality_2011}.
The de~Bruijn graph made from raw reads (i.e.\@ not error corrected) is likely larger and more complicated, making the assembly process more difficult and error prone.
In general, error correcting the reads leads to assemblies with longer contiguous sequences and fewer misassemblies~\citep{salzberg_gage:_2011}.

For Illumina sequencing, the base quality degrades toward the 3' ends of the reads, therefore one can trim the reads either by a fixed amount or based on the quality values reported by the sequencing machine.
Although these simple trimming schemes will reduce the number of erroneous bases, it still leaves many errors in the reads and needlessly discards a lot of valid sequence. 
Aggressive trimming can result in fragmented assemblies.

On the other hand, trimming can be an integral part of error correction.
The distribution of sequencing errors on the reads is complex and for some percentage of the reads, the sequence beyond a certain point contains too many errors to be corrected or, even worse, does not correspond to any sequence in the original genome.
It is important to trim those reads to avoid misassemblies~\citep{salzberg_gage:_2011}.

There are published error correctors that use a variety of techniques to detect and correct these erroneous bases~\citep{yang_survey_2012}.

Random sequencing errors result in spurious $k$-mers which, with high probability, occur only rarely in the reads.
A common approach to error correction is to eliminate rare $k$-mers from the read database by either correcting or trimming or discarding the reads containing those $k$-mers~\citep{pevzner_eulerian_2001,chaisson_short_2008,zhao_edar:_2010,shi_parallel_2010}.
Hence, one determines a threshold for eliminating from the reads each $k$-mer whose multiplicity is below the threshold.

It is possible to modify this approach.
For example, instead of counting the number of occurrences of a $k$-mer, Quake~\citep{kelley_quake:_2010} uses quality scores to create a weight for each occurrence and computes a weighted sum.
It tries to eliminate $k$-mers whose weighted sum is below a threshold.

Setting such a threshold works well when there is uniform coverage or, at least, when there is high coverage throughout the entire genome.
In practice, with second generation sequencing, parts of the genome will have low or zero coverage.
The method with a threshold will accurately correct the regions with high coverage and will convert low coverage regions to zero coverage regions.
Destroying these low coverage regions creates gaps in the assembly: where there is $k$-mer in the genome which is not in the error correct reads, one can expect a gap at that location in the assembly.

Our approach with \quorum is to eliminate the threshold and preserve low coverage regions when possible.
Our approach is to question a $k$-mer in a read when as one moves along the read, there is a sudden drop to low coverage, as described in Methods.

Some other approaches are not based on the multiplicities of $k$-mers in the reads.
For example, Coral~\citep{salmela_correcting_2011} and Echo~\citep{kao_echo:_2011} use multiple alignment of the reads and statistical models of sequencing to correct misaligned bases.
HiTec~\citep{ilie_hitec:_2011} uses a suffix array to find and correct potentially erroneous bases.
The last three error correctors mentioned only attempt to make base substitutions, while Quake and \quorum will also trim reads.

We are reporting on a new error correction procedure and software package, named \quorum (Quality Optimized Reads from the University of Maryland), that provides both error trimming and error correction.
It is targeted at improving genome assembly. 
\quorum works as a stand-alone program and is also a component of the assembler MaSuRCA~\citep{aleksey_zimin_masurca_2013}.

It is also fast (1 billion 100-base reads in a day using one core and scales linearly with multiple cores).
It can tackle the large data sets produced by today's high throughput sequencing machine.
\quorum only corrects substitution errors, not insertions and deletions.
It is well suited for correcting reads sequenced using Illumina technology~\citep{bentley_accurate_2008}, where the substitutions errors are the most common.

We evaluate the error correction and trimming skill of \quorum and compare it to  other published error correctors (HiTec, Echo, Coral and Quake), on  three genomes that have Illumina reads and have genome assemblies of finished quality.
We compare the error correctors by aligning the original reads and the error corrected reads to the finished sequence and measure the improvements.

\begin{methods}
\section{Methods}

We could define each $k$-mer as gtrueh if it is in the finished  genome and gfalseh otherwise.
Our error correction procedure aims at producing a set of error-corrected reads (1) minimizing the number of distinct false $k$-mers in the set of reads and (2) maximizing the number that are true.
These two goals must be balanced.
For example the more severely we truncate reads, (e.g., whenever we doubt the verity of a base of read and are unsure how to correct it), then we do better on (1) and worse on (2).
We aim to find the sweet spot that produces the best assembly.
Missing numerous true $k$-mers will yield an assembly that is either fragmented or erroneous.
Leaving too many false $k$-mers makes assembly extremely difficult.

Let $D$ be a data base of $k$-mers and their \emph{counts} (i.e. the number of occurrences in the set of reads).
For example, $D$ can be all $k$-mers in all reads, or alternatively, all $k$-mers that are in some sense high quality, based on quality scores.
Let define $\cnt(m)$,  the gcount of a $k$-mer $m$h, to be the number of occurrences $m$ in $D$.

\p{Starting to correct a read.}
We begin by creating a string $S$, initially consisting of the string of bases in the read.
We will possibly modify the string to obtain the corrected read.
Starting from the 5f end of a read, we choose as a starting point for error correction the first $k$-mer with a count of at least $3$.
We call it the \emph{anchor} $k$-mer.
If none exists, the read is discarded.
We accept the anchor as valid.
We proceed forward and backward from this $k$-mer.
We only describe the forward procedure below, that is, correcting while moving toward the 3f end.
The backward procedure is essentially identical.

At each iteration, after accepting a $k$-mer $m$ as valid, we shift one base in the forward direction, and evaluate the correctness of that base as follows.
Let $b$ denote the next base in $S$ beyond $m$.
Let $mf$ denote the $(k-1)$-suffix of $m$ and $mfx$ denote the $k$-mer consisting of $mf$ with base $x$ appended.
For example $mfb$ will denote the $k$-mer in $S$, which may differ from the corresponding $k$-mer in the read if a recent correction has been made.
 
\p{The $k$-mer count that guarantees acceptance without change.}
We choose a number of occurrences $Q$ so that if  $m'b$ has count $Q$ or greater, then we accept $b$ as valid.
Choose the smallest $Q$ so that in an ideal case the probability that a given false $k$-mer $\mu$ satisfies $\cnt(\mu) \ge Q$ is less than $10^{-6}$;
the ideal case is when errors are independent of each other, the read coverage is uniform and the genome has no repeats.
We also assume that the probability of an error that converts one base into another is independent of the two bases in question.
If, for example, the read coverage is $50$ and the error rate per base in the original reads is $1\%$, then the Poisson distribution implies $Q=5$.
In other words, under those conditions, if $k$-mer $m'b$ has a count greater or equal to $Q$, it is likely true

Whether or not the ideal case holds, we never change such a base $b$.
Lower count $k$-mers are examined in more detail.

\p{Candidate for correction}
We say that if the $k$-mer count of $m'b$ is less than $Q$, then the base $b$ is a \emph{candidate} (for correction).
For purposes of exposition we select $b$ to be the particular base $A$, and we determined the three counts of the $k$-mer $m'x$, where $x$ is $C, G, T$.
When $b = N$ in the above rules, we treat the count of $m'b$ as $0$.
There are a number of cases to consider, based on how many of these four counts are non-zero.
\begin{description*}
\item[(0) No non-zero count.] Then the corrected read is truncated at the current position.
\item[(1) Only one non-zero count.] Then set $b$ to the base with non-zero count, possibly $b$ itself.
Then we accept $m'b$ as valid and move on to the next base.
\end{description*}

\p{Two or more non-zero counts.}
Then, as described below, we attempt to eliminate choices in the hope that it will reduce to one of the above two cases.
For each $x$ where $\cnt(m'x) > 0$, we consider whether the $k$-mer $m'x$ has a one base extension with non-zero count.
Let $m''$ denotes the $k-2$-suffix of $m'$.
We ask if there is a base $y$ for which $\cnt(m''xy)$ is non-zero.
Specifically, we redefine the count of $m'x$ to be $\max_y\cnt(m''xy)$.
Then, if either all counts are zero or only one is non-zero, apply the rules in the above cases (0) and (1).

If two or more of the redefined counts are non-zero, we attempt to maintain \emph{continuity of coverage}.
We pick the base $x$ such that $\cnt(m'x)$ is the closest to $\cnt(m)$.
Finally, in case of a tie, we accept $m'b$ as valid (provided that $\cnt(m'b) > 0$, otherwise truncate at current position).

\p{Too many corrections.}
As we move along the read, we ``trust'' the corrections made so far.
In other words, if a correction was made to the string $S$ in the last $k-1$ bases, the $k$-mer $m'b$ matches the sequence $S$ and not of the original read.
There is a risk of rewriting a significant part of a read with sequence from another read, hence creating a \emph{chimeric} corrected read.
To avoid this risk, our procedure will truncate a read when too many corrections are made in a given window, by default $3$ corrections in a window of $10$ bases.
In such cases, we truncate so as to eliminate all $3$ corrections.

\p{Contaminant $k$-mers.}
Some reads may contain $k$-mers that do not belong to the genome but do belong to a known contaminant, such as a likely bacterial contaminant or adapter sequences.
Our procedure optionally takes as an input a list of $k$-mers in known contaminants.
\quorum truncates a read when encountering a contaminant $k$-mer, either before or after correction.
If an error correction produces a $k$ mer in the contaminant list, the read is truncated.
Such a $k$-mer will not appear in any of the corrected reads.

\p{Caveat.}
Above, we mention the possibility of using for the database $D$ either all $k$-mers in all reads ($D_{all}$) or all $k$-mers that are in some sense high quality ($D_{high}$).
Then we described the various options for changing a base using $D$ to determine which option is preferable.
In fact, \quorum uses both of those choices of $D$ simultaneously.
We do not review here all of the logical situations involving both databases.
As a guideline, when comparing two options, the high quality choice from the database $D_{high}$ always trumps the lower database $D_{all}$.

\end{methods}

\section{Discussion}
We evaluate the error correction software by error correcting the reads of three organisms, two bacterial genomes and a mammalian genome, for which a finished sequence is available: \species{Rhodobacter sphareoides} (rhodobacter)~\citep{mackenzie_home_2001}, \species{Staphylococcus aureus} (staphylococcus), and \species{Mus musculus} (mouse)~\citep{waterston_initial_2002}.
For the sake of simplicity, we only use chromosome 16 of the mouse genome which has \SI{98}{Mb} in finished sequence.
These genomes present different type of challenges for error correction.
The rhodobacter genome (\SI{4.6}{Mb} long) has a high GC content and is consequently difficult to sequence using Illumina technology.
The staphylococcus genome is \SI{2.9}{Mb} long.
The mouse chromosome is larger and has a more complex repeat structure.

We were able to successfully run all five error correctors on the bacterial data sets.  
Echo and HiTEC did not run to completion on the larger mouse data.
In addition, we implemented two simple programs that only trim the input reads.
trim20B trims $20$ bases from the 3' end of the reads, while trimQual5 trims the 3' end of a read at a base where the quality goes below or equal to $5$ and subsequently never goes above $5$.
When applicable, we also compare the results with making no correction at all, mentioned in the result tables as the ``none'' corrector.

\p{Metrics for effectiveness of error correction for genome assembly.}
A corrector can foul some of the reported metrics by very aggressively trimming the reads.
At the extreme, a corrector can trim every read to one base and, as a result, obtain only perfect reads.

A sequencing error in a read generates, with high probability, a collection of $k$-mers that occur only once in the reads.
Hence, trimming off $k$-mers that occur rarely in the reads may seem like a good idea.
However, because of occasional low coverage regions, some rare $k$-mers do represent actual sequence.
Eliminating these rare but real $k$-mers means that the very low coverage areas of the genome may wind up with no coverage by corrected reads, increasing the number of gaps in the assembly.

\p{False and missing $k$-mers.}
A $k$-mer is considered \emph{false} if it is present in the corrected reads and not in the finished sequence.
Note that a false $k$-mer present in multiple reads is counted as $1$ false $k$-mer.
Conversely, a $k$-mer is \emph{missing} if it is present in the finished sequence and not in the corrected reads.

\begin{table}[!t]
  \robustify\bfseries
  \processtable{Percent of false $31$-mers remaining and true $31$-mers missing in error corrected reads\label{table:k-mer-accuracy}}
  {\begin{tabular}{@{}
        l                               %
        S[detect-weight,table-format=3.1] %
        S[detect-weight,table-format=1.2] %
        S[detect-weight,table-format=3.1] %
        S[detect-weight,table-format=1.2] 
        S[detect-weight,table-format=3.0]  %
        S[detect-weight,table-format=1.2]
        @{}}
      \toprule
      Corrector & \multicolumn{2}{c}{Rhodobacter} & \multicolumn{2}{c}{Staphylococcus} & \multicolumn{2}{c}{Mouse}         \\
      \cmidrule(rl){2-3}                \cmidrule(rl){4-5}                   \cmidrule(l){6-7} %
      {}        & {False}      & {True}        & {False}      & {True}        & {False}    & {True}        \\
      {}        & {remain}     & {missing}     & {remain}     & {missing}     & {remain}   & {missing}     \\ 
      \midrule
      none      & 100          & 0.36          & 100          & 0.04          & 100        & 0.07          \\ 
      trim20B   & 55           & 0.39          & 64           & 0.09          & 50         & \winner{0.08} \\
      trimQual5 & 9            & 0.71          & 96           & 0.04          & 34         & 0.10          \\
      coral     & 69           & 0.38          & 56           & 0.13          & 52         & 0.22          \\
      echo      & 60           & \winner{0.36} & 55           & \winner{0.03} &            &               \\
      hitec     & 42           & 1.13          & 33           & 0.23          &            &               \\
      quake     & \winner{0.2} & 1.16          & 3.3          & 0.24          & 5          & 0.16          \\
      \quorum   & 0.4          & 0.40          & \winner{0.2} & 0.09          & \winner{2} & 0.11          \\
      \bottomrule
    \end{tabular}
  }{}
\end{table}

Table~\ref{table:k-mer-accuracy} is made by computing the symmetric difference between the $31$-mers in the finished sequence and in the corrected reads.
The ``False $31$-mers remaining'' is the number of false $31$-mers left in the corrected reads as a percentage of the false $31$-mers in the original reads.
By definition, the ``none'' error corrector has $100\%$ of its false $31$-mers remaining.
The numbers of true $k$-mers missing from the original reads are $16\,452$, $1\,047$, and $59\,322$ respectively for Rhodobacter, Staphylococcus, and Mouse.
The ``True $31$-mers missing'' is the number of missing $31$-mers as a percentage of the number of $31$-mers in the finished sequence.

We have found that low counts for both false and missing $k$-mers correlates well with better assembly quality.
Intuitively, having many false $k$-mers in a set of reads makes the creation of contigs more difficult for an assembler while missing many $k$-mers leads to a fragmented assembly.
This trend is partially supported by the results in Table~\ref{table:icontig}.

\quorum consistently has low counts for both the false and missing $k$-mer values while other correctors tend to have less balanced results: either the false or missing $k$-mer value is high.
For example, echo's staphylococcus corrected reads have fewer missing $k$-mers than the original reads, i.e.\@ it recovers some true $k$-mers which are not present in the original reads.
But it is not very aggressive in its correction and leaves many false $k$-mers.
On the other hand, quake usually leaves few false $k$-mers but is so aggressive in its trimming, that many true $k$-mers are missing from its corrected reads.

\begin{table}[!t]
  \robustify\bfseries
  \processtable{Idealized contig size statistics (in kb)\label{table:icontig}}
    {\begin{tabular}{@{}
        l %
        S[detect-weight,table-format=2.1]  %
        S[detect-weight,table-format=2.1]  %
        S[detect-weight,table-format=3.0]  %
        S[detect-weight,table-format=3.0]
        S[detect-weight,table-format=2.1]  %
        S[detect-weight,table-format=2.1]
        @{}}
      \toprule
      Corrector & \multicolumn{2}{c}{Rhodobacter} & \multicolumn{2}{c}{Staphylococcus} & \multicolumn{2}{c}{Mouse}    \\
      \cmidrule(rl){2-3}                \cmidrule(rl){4-5}                   \cmidrule(l){6-7} %
      {}        & {N50}         & {E-size}      & {N50}        & {E-size}     & {N50}         & {E-size}      \\
      \midrule                                                                                                       
      none      & 4.1           & 6.3           & 104          & 119          & 51.1          & 64.6          \\
      trim20B   & 7.9           & 10.8          & 64           & 80           & 59.5          & 74.0          \\
      trimQual5 & 6.0           & 7.9           & 98           & 101          & 65.5          & 84.5          \\
      coral     & 7.9           & 10.4          & 151          & 148          & 26.9          & 34.9          \\
      echo      & 10.4          & 15.1          & 162          & 217          &               &               \\
      hitec     & 9.3           & 12.9          & 65           & 83           &               &               \\
      quake     & 4.4           & 6.2           & 26           & 33           & 54.0          & 69.2          \\
      \quorum   & \winner{25.3} & \winner{30.6} & \winner{295} & \winner{268} & \winner{74.3} & \winner{92.5} \\
      \bottomrule
    \end{tabular}
  }{}
\end{table}

Table~\ref{table:icontig} gives an evaluation of how good the corrected reads might be for creating an assembly.
We error correct reads in order to improve the resulting assembly.
Important metrics of the quality of an assembly include the statistics on the length of the contigs generated, such as the ``N50'' size and the ``E-size''.
In general, the N$x$ size is defined as the contig size such that $x\%$ of the genome is contained in contigs of size N$x$ or larger.
For each base (i.e.\@ location) in the finished sequence compute the size of the contig it lies in.
The E-size is the average of these sizes, averaged over all the bases in the finished sequence.
In other words, the E-size is the expected contig size for a randomly chosen base in the genome.
It is computed as the sum of the square of the contig lengths over the genome length.

We estimate here an upper-bound on the N50 and E-size that one might expect, and these estimates are independent of particular assembly programs.
An \emph{idealized contig} consists of a segment of the genome that is covered by overlapping corrected reads, overlapping by at least $5$ bases.
These reads must match the finished sequence along their entire length with at least $98\%$ identity.
These estimates are upper-bounds in the sense that an assembler would probably require overlaps of more than $5$ bases and might require that multiple reads cover each base of a contig.
We note that reads with less than $98\%$ identity will likely be assembled in different contigs by the assembly program.
The N50 and E-size of idealized contigs are reported in Table~\ref{table:icontig}.

The correctors that do not trim output fewer reads that are error free on their entire length.
Hence, more reads failed to align to the finish sequence, resulting in gaps between the idealized contigs.
Conversely, Quake is more aggressive in its trimming and discards valuable sequence, again resulting in gaps between the idealized contigs.
The next table shows why \quorum reads produce bigger contigs.

\begin{table}[!t]
  \robustify\bfseries
  \processtable{Percentage of perfect reads and percentage of sequence contained in perfect reads.\label{table:perfect-reads}}
  {\begin{tabular}{@{}
        l                                  %
        S[detect-weight,table-format=2.1]  %
        S[detect-weight,table-format=2.1]  %
        S[detect-weight,table-format=2.1]  %
        S[detect-weight,table-format=2.1]
        S[detect-weight,table-format=2.1]  %
        S[detect-weight,table-format=2.1]
        @{}}
      \toprule
      Corrector & \multicolumn{2}{c}{Rhodobacter} & \multicolumn{2}{c}{Staphylococcus} & \multicolumn{2}{c}{Mouse}    \\
      \cmidrule(rl){2-3}                \cmidrule(rl){4-5}                   \cmidrule(l){6-7} %
      {}        & {Reads}       & {Sequence}    & {Reads}       & {Sequence}    & {Reads}       & {Sequence}    \\
      \midrule                                                                                                       
      none      & 20.9          & 20.9          & 33.1          & 33.1          & 48.3          & 48.3          \\
      trim20B   & 44.4          & 35.6          & 46.3          & 37.2          & 79.4          & 63.6          \\
      trimQual5 & 76.2          & 51.2          & 35.3          & 35.1          & 78.0          & 71.9          \\
      coral     & 58.2          & 58.2          & 73.5          & 73.5          & 80.9          & 80.9          \\
      echo      & 56.3          & 56.3          & 65.1          & 65.1          &               &               \\
      hitec     & 60.8          & 60.8          & 77.9          & 77.9          &               &               \\
      quake     & 80.5          & 58.9          & 68.2          & 57.6          & 88.8          & 81.3          \\
      \quorum   & \winner{89.4} & \winner{77.5} & \winner{84.4} & \winner{80.7} & \winner{91.6} & \winner{87.5} \\
      \bottomrule
    \end{tabular}
  }{}
\end{table}

Table~\ref{table:perfect-reads} reports the number of reads and the total sequence in perfect corrected reads.
All percentages are reported as a percentage of the original reads (number of original reads and sequence in original reads respectively).
A perfect read is defined as read having a full length error free alignment with the finished sequence.
\quorum consistently produces the largest number of perfect reads and the largest amount of sequence in perfect reads.
Although \quorum trims and discards some reads, the amount of perfect sequence is larger than that produced by the correctors that do not trim or discard any reads, such as Echo, HiTEC and Coral.
Moreover, the difference between the percentage in number of reads and percentage in sequence is due exclusively to trimming (the ratio is the average read length).
The non-trimming correctors have the same value for both column.
\quorum trims less aggressively than Quake, it produces longer read in average with a higher percentage of perfect reads.

\begin{table}[!t]
  \robustify\bfseries
  \processtable{Number of chimeric reads\label{table:chimeric}}
  {\begin{tabular}{@{}
        l                               %
        S[detect-weight,table-format=4] %
        S[detect-weight,table-format=4] %
        S[detect-weight,table-format=6] %
        @{}}
      \toprule
      {Corrector} & {Rhodobacter} & {Staphylococcus} & {Mouse}        \\
      \midrule                                                       
      none        & 2086          & 733              & 296772         \\
      trim20B     & 1655          & 597              & 250382         \\
      trimQual5   & 649           & 720              & 148439         \\
      coral       & 2047          & 1024             & 225176         \\
      echo        & 1809          & 704              &                \\
      hitec       & 6375          & 1200             &                \\
      quake       & \winner{61}   & \winner{518}     & \winner{62973} \\
      \quorum     & 236           & 786              & 81995          \\
      \bottomrule
    \end{tabular}
  }{}
\end{table}

Table~\ref{table:chimeric} reports the effect of error correction on chimeric reads.
We call a read \emph{chimeric} when it merges together sequences from two or more distant regions of the genome.
Such reads typically cannot be corrected, they have to be trimmed or discarded.
The Illumina technology generates few chimeric reads (usually less than $1\%$ of all the reads).
When doing very aggressive error correction, one runs the risk of creating new chimeric reads;
i.e.\@ sequence from a distant, possibly repeated, region of the genome is used to rewrite significant portion of a read.

Quake performs best at reducing the number of chimeric reads.

\section{Conclusion}
From the point of view of using error correctors as aids in the assembly of whole genome shotgun reads, the most revealing criterion is the size of the idealized contigs (Table~\ref{table:icontig}).
Overall, by most criteria in this paper, \quorum is the best error corrector for the purpose of genome assembly.
\quorum produces a larger proportion of error free reads (see Table~\ref{table:perfect-reads}) and its reads yield the largest idealized contig size (see Table~\ref{table:icontig}).
Chimeric reads can result in misassemblies of contigs and quake is best at eliminating chimeric reads.
An assembly program can detect for such errors by using mate pairs.
For such a misassembly one would expect that no mate pairs would straddle the chimeric join in the read.
The impact of any error corrector on the resulting assembly depends on the assembly program it is used with and on the genome assembled~\citep{magoc_gage-b:_2013}.

Despite our emphasis on genome assembly, \quorum is also being evaluated for cleaning reads for SNP finding and for transcriptome assembly.

\section*{Acknowledgement}
\paragraph{Funding\textcolon}
This project was supported by Agriculture and Food Research Initiative Competitive Grant no.\@ 2008-04049 and 2010-15739-01 from the USDA National Institute of Food and Agriculture and Grant R01HG002945 from the National Institutes of Health .

\bibliographystyle{natbib}
\bibliography{document-arxiv}
\end{document}